\documentclass[12pt]{article}

\usepackage{setspace}

\usepackage{amsmath}
\usepackage{graphicx}
\usepackage{natbib}
\usepackage{url} 
\usepackage{latexsym}
\usepackage{amsfonts}
\usepackage{amssymb}
\usepackage{psfrag}
\usepackage{graphicx}
\usepackage{calc, cancel}
\usepackage{graphics, graphicx, caption, subcaption}
\usepackage{latexsym, pgfplots, tikz, url,verbatim,multicol,enumerate,bm,listings,color}
\usepackage{multirow}
\usepackage[justification=centering]{caption}

\usepackage{wrapfig, sidecap,enumitem}
\usepackage{epsfig}
\usepackage{longtable}
\usepackage{hyperref}
\usepackage{graphicx}
\usepackage{natbib}
\usepackage{color}
\usepackage{amsgen,amsmath,amstext,amsbsy,amsopn,amssymb}
\usepackage{geometry}
\usepackage{subcaption}
\usepackage{caption}
\usepackage{wrapfig}
\usepackage{fancyhdr}
\usepackage{lastpage}
\usepackage{longtable}
\usepackage{enumitem}

\newcommand{\bB}{\mathbf{B}}
\newcommand{\bD}{\mathbf{D}}
\newcommand{\bI}{\mathbf{I}}
\newcommand{\bT}{\mathbf{T}}
\newcommand{\bu}{\mathbf{u}}
\newcommand{\bU}{\mathbf{U}}
\newcommand{\bV}{\mathbf{V}}
\newcommand{\by}{\mathbf{y}}
\newcommand{\bY}{\mathbf{Y}}
\newcommand{\bz}{\mathbf{z}}
\newcommand{\bZ}{\mathbf{Z}}
\newcommand{\bX}{\mathbf{X}}
\newcommand{\bzero}{\mathbf{0}}
\newcommand{\bepsilon}{\boldsymbol{\epsilon}}
\newcommand{\bmu}{\boldsymbol{\mu}}

\def\1v{\mathbf 1}
\def\0v{\mathbf 0}

\begin{document}
	\doublespacing
	
	\title{\bf A Family of Mixture Models for Biclustering}
	
	\author{Wangshu Tu\footnote{Department of Mathematical Sciences, Binghamton University, State University of New York, 4400 Vestal Parkway East, Binghamton, NY, USA 13902. e: wtu2@binghamton.edu} \and Sanjeena Subedi \footnote{Department of Mathematical Sciences, Binghamton University, State University of New York, 4400 Vestal Parkway East, Binghamton, NY, USA 13902. e: sdang@binghamton.edu}}

	\maketitle

\begin{abstract}
Biclustering is used for simultaneous clustering of the observations and variables when there is no group structure known {\it a priori}. It is being increasingly used in bioinformatics, text analytics, etc. Previously, biclustering has been introduced in a model-based clustering framework by utilizing a structure similar to a mixture of factor analyzers. In such models, observed variables $\bX$ are modelled using a latent variable $\bU$ that is assumed to be from $N(\mathbf{0}, \mathbf{I})$. Clustering of variables is introduced by imposing constraints on the entries of the factor loading matrix to be 0 and 1 that results in a block diagonal covariance matrices. However, this approach is overly restrictive as off-diagonal elements in the blocks of the covariance matrices can only be 1 which can lead to unsatisfactory model fit on complex data. Here, the latent variable $\bU$ is assumed to be from a $N(\mathbf{0}, \mathbf{T})$ where $\mathbf{T}$ is a diagonal matrix. This ensures that the off-diagonal terms in the block matrices within the covariance matrices are non-zero and not restricted to be 1. This leads to a superior model fit on complex data. A family of models are developed by imposing constraints on the components of the covariance matrix. For parameter estimation, an alternating expectation conditional maximization (AECM) algorithm is used. Finally, the proposed method is illustrated using simulated and real datasets.
\end{abstract}


\footnotetext{\textbf{Abbreviations:} AECM, alternating expectation conditional maximization; ALL, acute lymphoblastic leukemia; AML, acute myeloid leukemia; ARI, adjusted Rand index; BIC, Bayesian information criteria; EM, expectation-maximization; MFA, mixtures of factor analyzers; FDR, false discovery rate}

\section{Introduction}
Cluster analysis, also known as unsupervised classification, assigns observations into clusters or groups without any prior information on the group labels of any of the observations. It differs from supervised classification where training data with known labels are used to build models with the aim of classifying observations with no labels. In many situations, labels for all observations are not available in advance or are missing. In such cases, the observations are assigned to groups (or clusters) based on some measure of similarity (e.g., distance) \citep{review2017}. Using a similarity measure, the goal in clustering is to identify subgroups in a heterogeneous population such that individuals within a subpopulations are more homogenous compared to the entire population. Cluster analysis has been widely used to find hidden structures in  many fields such as bioinformatics for clustering genes \citep{jiang2004,mcnicholas2012}, image analysis \citep{houdard2018,gonzales2006}, market research for market segmentation \citep{saunders1980}, etc. Clustering algorithms can be broadly divided into hierarchical clustering approaches and partition-based clustering approaches. Hierarchical clustering \citep{ward1963,johnson1967} creates clusters of data using a tree like structure either by progressive fusion of clusters (i.e., agglomerative hierarchical clustering) or divisions of clusters (i.e., divisive hierarchical clustering). 
Partition-based approach include non-parametric approaches such as $k$-means \citep{zz} and parametric approaches such as model-based clustering. $k$-means partitions a data set into $k$ distinct, non-overlapping clusters using a predefined criteria. However, $k$-means and other similar approaches are highly dependent on starting values, correlation among variables are not taken into account for multivariate data, and can be sensitive to outliers \citep{sisodia2012}. Model-based clustering algorithms utilize finite mixture models and provide a probabilistic framework for clustering data. 
Such models assume that data comes from a finite collection of subpopulations or components where each subpopulation can be represented by a distribution function depending on the nature of the data. In particular, a $K$-component finite mixture density can be written as $$f(\by_i|\boldsymbol{\vartheta})=\sum_{k=1}^{K}\pi_kf_k(\by_i|\boldsymbol{\theta}_k),$$ where $\pi_k>0$ is the mixing portion such that $\sum_{k=1}^{K}\pi_k=1$, $f_k(\by_i|\boldsymbol{\theta}_k)$ is the density function of each component, and $\boldsymbol{\vartheta}=(\boldsymbol{\theta}_1,\boldsymbol{\theta}_2,\ldots,\boldsymbol{\theta}_K)$ represents the model parameters. In the last three decades, there has been an explosion in model-based approaches for clustering different types of data \citep{banfield1993,fraley2002,subedi2014,franczak14,dang2015,melnykov18,silva2019,subedi2020}.

Traditional clustering algorithms, here referred to as one-way clustering methods, aim to group observations based on similarities across all variables at the same time. This can be too restrictive as observations may be similar under some variables, but different for others \citep{Pad2017}. This limitation motivated the development of biclustering algorithms that simultaneously cluster both rows and columns, i.e., partitioning a data matrix into small homogeneous blocks \citep{mirkin1996}. The idea of biclustering was first introduced by \cite{hartigan-direct-clustering-data-1972} which proposed a partition based algorithm to find constant biclusters in a data matrix. \cite{Cheng2000} proposed another approach for biclustering that aimed to find homogeneous submatrices using a similarity score through iterative addition/deletion of rows/columns. A similar approach was also proposed by \cite{yang2002}. In \cite{kluger03spectralbiclustering}, the authors developed a biclustering that utilized a singular value decomposition of the data matrix to find biclusters. While this approach is computationally efficient compared to the previous approaches, the algorithm and interpretations of the biclusters are reliant on the choice of normalization. Authors in \cite{bendor2002}, \cite{murali2003}, and \cite{liuwang2003} focused on finding a coherent trend across the rows/columns of the data matrix regardless of their exact values rather than trying to find blocks with similar values.  In \cite{SAMBA}, authors introduced a biclustering method based on graph theory. This approach converts the rows and columns into a bipartite graph and tries to find the densest subgraphs in a bipartite graph. However, these approaches were often computationally intensive and there was a lack of statistical model on which inferences can be made.

In \cite{govaert2008}, authors introduced a model-based co-clustering algorithm using a latent block model for binary data by introducing an additional latent variable that are column membership indicators.  \cite{cocluster} extended this approach for contingency tables. A similar framework using a mixtures of univariate Gaussian distributions for biclustering continuous data was utilized by \cite{blockcluster}. Alternatively, \cite{Martella2008BiclusteringOG} proposed a model-based biclustering framework based on the latent factor analyzer structure. A factor analyzer model \citep{Spearman1904,Bartlett1953} assumes that an observed high dimensional variable $\bY$ can be modelled using a much smaller dimensional latent variable $\bU$. Incorporating this factor analyzer structure in the mixtures of Gaussian distribution, mixtures of factor analyzer have been developed by \cite{ghahramani1996,tipping1999,mclachlan2000}. Since then, mixtures of factor analyzers have been widely used for various data types  \citep{andrews2011,subedi2013,murray2014,subedi2015,lin2016,tortora2016}.  \cite{Martella2008BiclusteringOG}  replaced the factor loading matrix by a binary and row stochastic matrix and imposed constraints on the components of the covariance matrices resulting in a family of four models for model-based biclustering. In \cite{wong2017two}, the authors further imposed additional constraints on the components of the covariance matrices and the number of latent factors resulting in a family of eight models. However, one major limitation with both \cite{Martella2008BiclusteringOG}  and \cite{wong2017two} is that these models can only recover a restrictive covariance structure such that the off-diagonal elements in the block structure of the covariance matrices are restricted to be 1.

In this paper, we modify the assumptions for the latent factors in the factor analyzer structure  used by \cite{Martella2008BiclusteringOG} and \cite{wong2017two} to capture a wider range of covariance structures. This modification allows for more flexibility in the off-diagonal elements of the block structure of the covariance matrix. Furthermore, a family of parsimonious models is presented. The paper is organized as follows. Details of the generalization are provided in Section \ref{method} with details on parameter estimation and model selection. In Section \ref{result}, we show that these extensions allows for better recovery of the underlying group structure and can recover the sparsity in the covariance matrix through simulation studies and real data analyses. The paper concludes with a discussion and future directions in Section \ref{conc}.

\section{Methodology} \label{method}
\subsection{Factor analyzers based biclustering}\label{FAbiclust}
In the factor analysis model \citep{Spearman1904,Bartlett1953}, a $p$-dimensional variable $\bY_i$ can be written as
\[ \bY_i= \bmu+\bV\bU_i+\bepsilon_i, \quad (i=1,\ldots,n)\]
where $\bU_i\sim N_q(\bzero,\bI_q)$ is a $q$-dimensional ($q \ll p$) vector of latent factors, $\bepsilon_i \sim N_p(\bzero,\bD)$ where $\bD$ is a diagonal matrix and $\bepsilon_i$ is independent of $\bU_i$, $\bV$ is a $p\times q$ matrix of factor loadings, and $\bmu$ is a $p$-dimensional mean vector. Then,
\[ \bY_i\sim N_p(\bmu,\bV\bV^T+\bD),\]
and, conditional on $\bU$,
\[\bY_i\mid \bu_i \sim N_p(\bmu+\bV\bu_i,\bD).\]

In the mixture of factor analyzer models with $K$ components \citep{Ghahramani97theem,McLachlan00mixturesof,mcnicholas2008}, the $p$-dimensional variable $\bY_i$ can be modeled as
\[ \bY_{i} = \boldsymbol{\mu}_k+\mathbf{V}_k\bU_{ik}+\epsilon_{ik}, \quad \text{with prob}~\pi_k~(k=1,\ldots,K; i=1,\ldots,n),\]
where $\bU_{ik}\sim N(\bzero,\bI_{q_k})$ is a $q_k$ dimensional vector of latent factors in the $k^{th}$ component, $\boldsymbol{\mu}_k$ is the mean of the $k^{th}$ component, $\bV_k$ is $p\times q_k$ matrix of factor loadings of the $k^{th}$ component, and $\bI_{q_k}$ is an identity matrix of size $q_k$. \cite{Ghahramani97theem,McLachlan00mixturesof}, and \cite{mcnicholas2008} assume the same number of latent variables for all $K$ components (i.e., $q_1=\ldots=q_K=q$). In \cite{Martella2008BiclusteringOG} and \cite{wong2017two}, authors proposed a family of models by replacing the factor loading matrix $\bV$ with a binary row-stochastic matrix $\bB$. This $p\times q_k$ dimensional matrix $\bB$ can be regarded as cluster membership indicator matrix for the variable clusters (i.e. column clusters) such that $\bB[i,j]=1$ if the $i^{th}$ variable belongs to $j^{th}$ column clusters and $\bB[i,k]=0$ for all $k\neq j$. Under their framework, all clusters have the same number of latent variables, and fixed covariance($\bI$) for latent variables. By constraining the number and covariance of latent variables, the cluster covariance structure is very limited. The correlation
between variables will depend on $\bD$ only. For complex real data, this is overly restrictive as different clusters could have different number of latent variable, and covariance of the latent variable doesn't have to be $\bI$. To overcome the limitations, we propose a modified MFA model and extend it for biclustering. 

\subsection{Modified MFA and its extension for biclustering}		
Here, we utilize a modified the factor analyzer structure such that the $p$-dimensional variable $\bY_i$ can be modeled as
\[ \bY_{i} = \boldsymbol{\mu}_k+\bV_k\bU_{ik}+\epsilon_{ik}, \quad \text{with prob}~\pi_k~(k=1,\ldots,K; i=1,\ldots,n),\]
where we assume $\bU_{ik}\sim N(\bzero,\bT_{q_k})$, where $\bT_{q_k}$ is a diagonal matrix with entries $\{t_1, t_2,\cdots, t_{q_k}\}$. Additionally, we allow different clusters to have different number of latent variables. Hence, 
\[ \bY_i\sim N_p(\bmu_k,\bV_k\bT_{q_k}\bV_k^T+\bD_k),\]
and, conditional on $\bU_{ik}$,
\[\bY_i\mid \bu_{ik} \sim N_p(\bmu_k+\bV_k\bu_{ik},\bD_k).\]
In order to do biclustering, similar to  \cite{Martella2008BiclusteringOG}, we replace loading matrix $\bV_k$ by a sparsity matrix $\bB_K$ with entries $\bB_k[i,j]=1$ if $i^{th}$ variable belongs to $j^{th}$ group, 0 otherwise. 
Under this assumption, we are clustering the variables (i.e., columns) according to the underlying latent factors as each variable can only be represented by one factor and variables represented by the same factors are clustered together. In \cite{Martella2008BiclusteringOG}, the authors assumed $\bT_{q_k}=\bI_q$ and as stated in Section \ref{FAbiclust}, this imposes a stricter restriction on the structure of the component-specific of covariance matrices. This restriction not only influences recovering of the true component specific covariance but also affects the clustering of the observations (i.e., rows). By assuming $\bT_{q_k}$ to be a diagonal matrix with entries $\{t_1, t_2,\cdots, t_{q_k}\}$, the component specific covariance matrix becomes a block-diagonal matrix and within the block matrix, the off-diagonal elements are not restricted to 1. For illustration, suppose we have
\[\bB_k=
\left[\begin{matrix}
1 & 0 & 0\\
1 & 0 & 0\\
1 & 0 & 0\\
0 & 0 &1 \\
0 & 0 & 1\\
0 & 1 &0\\
0 & 1 &0
\end{matrix}\right]_{7\times 3},
\] $\bT_{q_k}=\text{diag}(t_1, t_2,\cdots, t_{3})$, and $\bD_k=\text{diag}(d_1,d_2,\ldots,d_7)$, the resulting component specific covariance matrix becomes
\[
\bB_k\bT_{q_k}\bB_k^T+\bD_k=\left[\begin{matrix}
t_1+d_1&t_1&t_1&0&0&0&0\\
t_1&t_1+d_2&t_1&0&0&0&0\\
t_1&t_1&t_1+d_3&0&0&0&0\\
0&0&0&t_2+d_4&t_2&0&0\\
0&0&0&t_2&t_2+d_5&0&0\\
0&0&0&0&0&t_3+d_6&t_3\\
0&0&0&0&0&t_3&t_3+d_7
\end{matrix}\right].
\]

Therefore, with different combination of $t$s and $d$'s, each block in the block-diagonal covariance matrix can capture:
\begin{itemize}[itemsep=-3pt,topsep=0pt]
	\item[-] large variance, low correlation;
	\item[-] small variance, high correlation;
	\item[-] large variance, high correlation; and
	\item[-] small variance, low correlation.
\end{itemize}
Recall that in \cite{Martella2008BiclusteringOG}  and in \cite{wong2017two}, $t$s are restricted to 1 and therefore, the model only allows for large variance and low correlation or small variance and high correlation. Additionally, \cite{wong2017two} imposed further restriction that all components must have the same number of latent factors (i.e., $q_1=q_2=\ldots=q_K=q$).

\subsection{Parameter Estimation}\label{PE}
Parameter estimation for mixture models is typically done using an expectation-maximization (EM) algorithm \citep{DEMP1977}. This is an iterative approach when the data are incomplete or are treated as incomplete. It involves two main steps: an expectation step (E-step) where the expected value of the complete-data log-likelihood is computed using current
parameter estimates, and a maximization step (M-step) where the expected value of the complete-data log-likelihood is then maximized with respect to the model parameters. The E- and M-steps are iterated until convergence. Herein, we utilize an alternating expectation conditional maximization(AECM) algorithm \cite{AECM}, which is an extension of the EM algorithm that uses different specifications of missing data at each stage/cycle and the maximization step is replaced by a series of conditional maximization steps. Here, the observed data $\bY_1,\ldots,\bY_n$ are viewed as incomplete data and the missing data arises from two sources: the unobserved latent factor $\bU_1,\ldots,\bU_n$ and the component indicator variable $\bZ_1,\ldots,\bZ_n$ where \[
z_{ik}=\begin{cases}
1 & \text{if observation}~ i\in \text{$k^{th}$ group}\\
0 & \text{otherwise}.
\end{cases}
\]

In first cycle, we treat $z_{ik}$ as the missing data. Hence, the complete data log-likelihood is 
\begin{equation*}\label{eq1}
\begin{split}
l_{1}(\bY,\bZ) &=\sum_{i=1}^{n}\log f(\by_i,\bz_{i})=\sum_{i=1}^{n}\log\prod_{k=1}^{K}\left\{\pi_kf_k(\by_i;\boldsymbol{\mu}_k, \bB_k\bT_{q_k}\bB_k^T+\bD_k)\right\}^{z_{ik}}\\
&=\sum_{i=1}^{n}\sum_{k=1}^{K}z_{ik}\left\{\log(\pi_k)+\log f_k(\by_i;\bmu_k, \bB_k\bT_{q_k}\bB_k^T+\bD_k)\right\}.
\end{split}
\end{equation*}
In the E-step, we compute the expected value of the complete data log-likelihood where the unknown memberships are replaced by their conditional expected values:
\[\hat{z}_{ik}=E(Z_{ik}|\by)=\frac{\hat{\pi}_kf_k(\hat{\boldsymbol{\mu}}_k, \hat{\bB}_k\hat{\bT}_{q_k}\hat{\bB}_k^T+\hat{\bD}_k)}{\sum_{k=1}^{K}\hat{\pi}_kf_k(\hat{\boldsymbol{\mu}}_k, \hat{\bB}_k\hat{\bT}_{q_k}\hat{\bB}_k^T+\hat{\bD}_k)}.\]
Therefore, the expected complete data log-likelihood becomes
\begin{align*}
Q_1(\boldsymbol{\mu}_k, \pi_k)&=\sum_{k=1}^{K}n_k\log(\pi_k)-\frac{np}{2}\log(2\pi)-\frac{1}{2}\sum_{k=1}^{K}n_k\log|\bB_k\bT_{q_k}\bB_k^T+\bD_k|\\
&-\frac{1}{2} \sum_{k=1}^K n_k\text{tr}\left\{\mathbf{S}_k(\bB_k\bT_{q_k}\bB_k^T+\bD_k)^{-1}\right\},
\end{align*}
where $n_k=\sum_{i=1}^{n}\hat{z}_{ik}$ and $\mathbf{S}_k=\frac{\sum_{i=1}^{n}\hat{z}_{ik}(\by_i-\hat{\boldsymbol{\mu}}_k)(\by_i-\hat{\boldsymbol{\mu}}_k)^T}{n_k}$. In the M-step, maximizing the expected complete data log-likelihood with respect to $\pi_k$ and $\boldsymbol{\mu}_k$ yields
\[
\hat{\pi}_k=\frac{n_k}{n},
\]
\[
\hat{\boldsymbol{\mu}}_k=\frac{\sum_{i=1}^{n}\hat{z}_{ik}\by_i}{n_k}.
\]

In the second cycle, we consider both $\bZ$ and $\bU$ as missing and the complete data log-likelihood in this cycle has the following form: 
\begin{equation*}
\begin{split}
l_2(\bY,\bU,\bZ)&=\sum_{i=1}^{n}\log f(\by_i,\bu_{i}, \bz_{i})=\sum_{i=1}^{n}\log \prod_{k=1}^{K}\left\{\pi_kf_k(\by_i\mid\bu_i;\boldsymbol{\mu}_k+\bB_k\bu_{ik}, \bD_k)f_k(\bu_i; \bzero,\bT_{q_k})\right\}^{z_{ik}}\\
&=\sum_{i=1}^{n}\sum_{k=1}^{K}z_{ik}\{\log\pi_k+\log f_k(\by_i\mid\bu_i;\boldsymbol{\mu}_k+\bB_k\bu_{ik}, \bD_k)+\log f_k(\bu_i; \bzero, \bT_{q_k})\}\\
&=C+\sum_{k=1}^{K}\left[\sum_{i=1}^nz_{ik} \left(\log \pi_k+\frac{1}{2}\log|\bD_k^{-1}|+\frac{1}{2}\log|\bT_{q_k}^{-1}|\right)-\frac{1}{2}\text{tr}\left(\bT_{q_k}^{-1}\sum_{i=1}^nz_{ik}\bu_{ik}\bu_{ik}^T\right)\right.\\
&-\frac{1}{2}\text{tr}\left\{\bD_k^{-1}\sum_{i=1}^nz_{ik}(\by_i-\bmu_k)(\by_i-\bmu_k)^T\right\}+\sum_{i=1}^{n}z_{ik}(\by_i-\boldsymbol{\mu}_k)^T\bD_k^{-1}\bB_k\bu_{ik}\\
&\left.-\frac{1}{2}\text{tr}\left\{\bB_k^T\bD_k^{-1}\bB_k\sum_{i=1}^{n}z_{ik}\bu_{ik}\bu_{ik}^T\right\}\right],
\end{split}
\end{equation*}
where C is some value that does not depend on $\bB_k, \bD_k, \bT_{q_k}, \bu_{ik}, \bz_i$, and $\pi_k$. 

Therefore, to compute the expected complete data log-likelihood, we must calculate the following expectations: $E(Z_{ik}\mid\by_i)$, $E(Z_{ik}\bU_{ik}\mid\by_i)$, and $E(Z_{ik}\bU_{ik}\bU_{ik}^T\mid\by_i)$.
We have
\[
\left[\begin{matrix}
\by_i\\
\bu_{ik}
\end{matrix}
\right]|z_{ik}\sim MVN\left[\begin{matrix}
\left(\begin{matrix}
\boldsymbol{\mu}_k\\
0
\end{matrix}\right), & \left(\begin{matrix}
\bB_k\bT_{q_k}\bB_k^T+\bD_k & \bB_k\bT_{q_k}\\
\bT_{q_k}\bB_k^T & \bT_{q_k}
\end{matrix}\right)
\end{matrix}\right].
\]
Therefore, 
\begin{align*}
E(\bU_{ik}|\by_i, z_{ik}=1)&=\bT_{q_k}\bB_k^T(\bB_k\bT_{q_k}\bB_k^T+\bD_k)^{-1}(\by_i-\boldsymbol{\mu}_k):=\hat{\bu}_{ik},\\
E(\bU_{ik}\bU_{ik}^T|\by_i,z_{ik}=1)&=\bT_{q_k}-\bT_{q_k}\bB_k^T(\bB_k\bT_{q_k}\bB_k^T+\bD_k)^{-1}\bB_k\bT_{q_k}+\hat{\bu}_{ik}\hat{\bu}_{ik}^T:=\boldsymbol{\theta}_k n_k.
\end{align*}

Then, the expectation of the complete data log-likelihood $Q_2$ can be written as:
\begin{equation}\label{UUUQ2}
\begin{split}
Q_2(\bB_k, \bD_k, \bT_{q_k})=&C_2+\sum_{k=1}^{K}\frac{n_k}{2}\left[\log|\bD_k^{-1}|+\log|\bT_{q_k}^{-1}|-tr\{\bD_k^{-1}\boldsymbol{S}_k\}-tr\{\bT_{q_k}^{-1}\boldsymbol{\theta}_k\}\right.\\
&\left.+2\sum_{i=1}^{n}\hat{z}_{ik}(\by_i-\hat{\boldsymbol{\mu}}_k)^T\bD_k^{-1}\bB_k\hat{\bu}_{ik}-tr\{\bD_k^{-1}\bB_k\boldsymbol{\theta}_k\bB_k^T\}\right],
\end{split}
\end{equation}
where $C_2$ stands for terms that are independent of $\bB_k, \bD_k, \bT_{q_k}, \bu_{ik}$, and $n_k=\sum_{i=1}^n\hat{z}_{ik}$. 
In the M-step, maximizing the expected value of the complete data log-likelihood with respect to $\bD_k$ and $\bT_{q_k}$ yields
\begin{align*}
\hat{\bD}^{(t+1)}_k&=\text{diag}\{\mathbf{S}_k-2\hat{\bB}_k\hat{\bT}_{q_k}\hat{\bB}_k^T(\hat{\bB}_k\hat{\bT}_{q_k}\hat{\bB}_k^T+\hat{\bD}_k^{(t)})^{-1}\mathbf{S}_k+\hat{\bB}_k\boldsymbol{\theta}_k\hat{\bB}_k^T\},\\
\hat{\bT}^{(t+1)}_{q_k}&=\text{diag}\left(\boldsymbol{\theta}_{ik}\right)=\text{diag}\left(\hat{\bT}^{(t)}_{q_k}-\hat{\bT}^{(t)}_{q_k}\hat{\bB}_k^T(\hat{\bB}_k\hat{\bT}^{(t)}_{q_k}\hat{\bB}_k^T+\hat{\bD}_k)^{-1}\hat{\bB}_k\hat{\bT}^{(t)}_{q_k}+\frac{\sum_{i=1}^n\hat{z}_{ik}\hat{\bu}_{ik}\hat{\bu}_{ik}^T}{n_k}\right).
\end{align*}

When estimating $\bB_k$, we choose $\bB_k[i,j]=1$ when $\bB_k$ maximizes $Q_2$ with a constraint that $\sum_{j=1}^{q_k}\bB_k[i,j]=1$ for all $k$.\\

Overall, the AECM algorithm consists of the following steps:
\begin{enumerate}
	\item [1] Determine the number of clusters: $K$ and $q_k$, then give initial guesses for $\bB_k, \bD_k, \bT_{q_k}$ and $z_{ik}$. 
	
	\item [2] First cycle:
	\begin{itemize}
		\item [(a)] E-step: update $z_{ik}$
		\item [(b)] CM-step: update $\pi_k, \boldsymbol{\mu}_k$
	\end{itemize}
	\item [3] Second cycle:
	\begin{itemize}
		\item [(a)] E-step: update $z_{ik}$ again and $\bu_{ik}$. 
		\item [(b)] CM-step: update $\mathbf{S}_k, \bD_k, \bT_{q_k}, \bB_k$
	\end{itemize}
	\item [4] Check for convergence. If converged, stop, otherwise go to step 2. 
\end{enumerate}

\subsection{A family of models}
To introduce parsimony, constraints can be imposed on the components of the covariance matrices $\bB_k$, $\bT_{q_k}$ and $\bD_k$ that results in a family of 16 different models with varying number of parameters (see Table \ref{family}). Here, ``U" stands for unconstrained, ``C" stands for constrained. This allows for a flexible set of models with covariance structures ranging from extremely constrained to completely unrestricted. Note that the biclustering model by \cite{Martella2008BiclusteringOG} can be recovered by imposing a constraint such that $q_1=q_2=\ldots=q_K=q$ and $\bT_{q_k}=\bI_q$. Details on the parameter estimates for the entire family is provided in the Appendix \ref{appendix: family}.
\begin{table}[!ht]
	\scriptsize
	\centering
	\caption{Parsimonious family of models obtained by imposition of constraints on $\bB_k$, $\bT_{q_k}$ and $\bD_k$.}\label{family}
	\begin{tabular}{@{\extracolsep{5pt}}lccccc}
		\\[-1.8ex]\hline
		\hline \\[-1.8ex]
		\multicolumn{1}{c}{Model} &  \multicolumn{1}{c}{$\bB_k$}&\multicolumn{1}{c}{$\bT_{q_k}$}&\multicolumn{2}{c}{$\bD_k$}&\multicolumn{1}{c}{Total number of parameters}\\
		\cline{3-3}\cline{4-5}
		&$\bB_k=\bB$&$\bT_{q_k}=\bT$&$\bD_k=\bD$& $\bD_k=d_k \bI$&\\
		\hline\\[-1.8ex]
		UUUU & U&U&U&U&p*K+$\sum_{k=1}^{K}q_k$+p*K+K-1+p*K\\
		UUUC&U&U&U&C&p*K+$\sum_{k=1}^{K}q_k$+K+K-1+p*K\\
		UUCU&U&U&C&U&p*K+$\sum_{k=1}^{K}q_k$+p+K-1+p*K\\
		UUCC&U&U&C&C&p*K+$\sum_{k=1}^{K}q_k$+1+K-1+p*K\\
		UCUU & U&C&U&U&p*K+$q_k$+p*K+K-1+p*K\\
		UCUC&U&C&U&C&p*K+$q_k$+K+K-1+p*K\\
		UCCU&U&C&C&U&p*K+$q_k$+p+K-1+p*K\\
		UCCC&U&C&C&C&p*K+$q_k$+1+K-1+p*K\\
		CUUU & C&U&U&U&p+$\sum_{k=1}^{K}q_k$+p*K+K-1+p*K\\
		CUUC&C&U&U&C&p+$\sum_{k=1}^{K}q_k$+K+K-1+p*K\\
		CUCU&C&U&C&U&p+$\sum_{k=1}^{K}q_k$+p+K-1+p*K\\
		CUCC&C&U&C&C&p+$\sum_{k=1}^{K}q_k$+1+K-1+p*K\\
		CCUU& C&C&U&U&p+$q_k$+p*K+K-1+p*K\\
		CCUC&C&C&U&C&p+$q_k$+K+K-1+p*K\\
		CCCU&C&C&C&U&p+$q_k$+p+K-1+p*K\\
		CCCC&C&C&C&C&p+$q_k$+1+K-1+p*K\\
		\hline \\[-1.8ex]
	\end{tabular}
\end{table}


\subsection{Initialization}
Mixture models are known to be heavily dependent on model initialization. 
Here, the initial values are chosen as following:
\begin{enumerate}
	\item$z_{ik}^{(ini)}$: The row cluster membership indicator variable $z_{ik}$ can be initialized by performing an initial partition using $k$-means, hierarchical clustering, random partitioning, or fitting a traditional mixture model-based clustering. Here, we chose the initial partitioning obtained via Gaussian mixture models available using the {\sf R} package ``mclust"\cite{mclust}. 
	\item $\bD_k^{(ini)}, \bT_{q_k}^{(ini)}$: Similar to \cite{mcnicholas2008}, we estimate the sample covariance matrix $\mathbf{S}_k$ for each group and then use the first $q_k$ principle components as $\bV_k\boldsymbol{\lambda}_k \bV_k^T$ where $\bV_k$ is the first $q_k$ principle loading matrix and $\boldsymbol{\lambda}_k$ is the first $q_k$ variance of principle components.  
	\[
	\bD_k^{(ini)}=\text{diag}\{\mathbf{S}_k-\bV_k\bT_k \bV_k^T\}, \text{and }  \bT_{q_k}^{(ini)}=\boldsymbol{\lambda}_k.
	\]
	\item $\bB_k^{(ini)}$: Similar to Step 2, but we use a scaled version of PCA to get the loading matrix $\mathbf{L}_k$. Then for each row i, let $\mathbf{L}_k[i,j]=1$ if $\mathbf{L}_k[i,j]=max_h\{\mathbf{L}_k[i,h]\}$, 0 otherwise. 
\end{enumerate}

\subsection{Convergence, model selection and label switching}
For assessing convergence, Aitken's convergence criteria \citep{aitken26} is used. 
The Aitken's acceleration at iteration $t$ is defined as:
\[
a^{(t)}=\frac{l^{(t+1)}-l^{(t)}}{l^{(t)}-l^{(t-1)}},
\]
where $l^{(t+1)}$ stands for the log-likelihood values at $t+1$ iteration. Then the asymptotic estimate for log-likelihood at iteration $t+1$ is:
\[
l_{\infty}^{(t+1)}=l^{(t)}+\frac{l^{(t+1)}-l^{(t)}}{1-a^{(t)}}.
\]
The AECM can be considered converged when
\[
|l_{\infty}^{(t+1)}-l_{\infty}^{(t)}|<\epsilon,
\]
where $\epsilon$ is a small number \citep{bohning94}. Here, we choose $\epsilon=10^{-2}$.

In the clustering context, the true number of components are unknown. The EM algorithm or its variants are typically run for a range of possible number of clusters and model selection is done {\it a posteriori} using a model selection criteria. Here, the number of latent factors $q_k$ is also unknown. Therefore, the AECM algorithm is run for all possible combinations of the number of clusters and number of latent variables, and the best model is chosen using the Bayesian Information Criterion \citep[BIC;][]{schwarz197801}. Mathematically, \[
BIC=2~L(\by,\boldsymbol{\hat{\vartheta}})- m \log(n),
\]
\noindent where $L(\by,\boldsymbol{\hat{\vartheta}})$ is the log-likelihood evaluated using the estimated
parameters, $m$ is the number of free parameters, and $n$ is the
number of observations. For performance evaluation, we use the adjusted Rand index \citep[ARI;][]{Hubert85comparingpartitions} when the true labels are known. The ARI is 1 for perfect agreement while the expected value of ARI is 0 under random classification. In one-way clustering, label switching refers to the invariance of the likelihood when the mixture component labels are relabelled \citep{Stephens00dealingwith} and it is typically dealt with imposition of identifiability constraints on the model parameters. In biclustering, both the row and column memberships could be relabelled. The identifiability of the row membership is ensured by imposing constraints on the mixing proportions such that $\pi_1\geq\pi_2\ldots\geq\ldots\geq\pi_K$. For column clusters, interchanging the columns of $\bB_k$ doesn't change column cluster membership, however, the associated diagonal elements of the matrix $\bT_{q_k}$ as well as the error matrix $\mathbf{D}_k$ needs to be permuted in order to recover the covariance matrix correctly. Failure to do so may trap or decrease overall likelihood.  In order to overcome this issue, if overall likelihood decreased at $(t+1)^{th}$ iteration, we assign $\bB_k^{(t+1)}=\bB_k^{(t)}$, otherwise $\bB_k^{(t+1)}=\bB_k^{(t+1)}$.


\section{Results}\label{result}
We did two sets of simulation studies. For each simulation study, we generate one hundred eight-dimensional datasets, each of size $n = 1000$ and ran all of our sixteen proposed models for $K = 1\ldots4$ and $q_k = 1\ldots4$. We also ran the unconstrained UUU model by \cite{wong2017two}. Note that this model can be obtained as a special case of our UUUU model by constraining $q_1=q_2=\ldots=q_K=q$ and $\bT_{q_k}=\bI$. For each dataset, the model with the highest BIC is chosen a posteriori among all the models including the model by \cite{wong2017two}.
\subsection{Simulation Study 1}
\begin{figure}[!h]\centering
	\includegraphics[width=5.5in,height=3.5in]{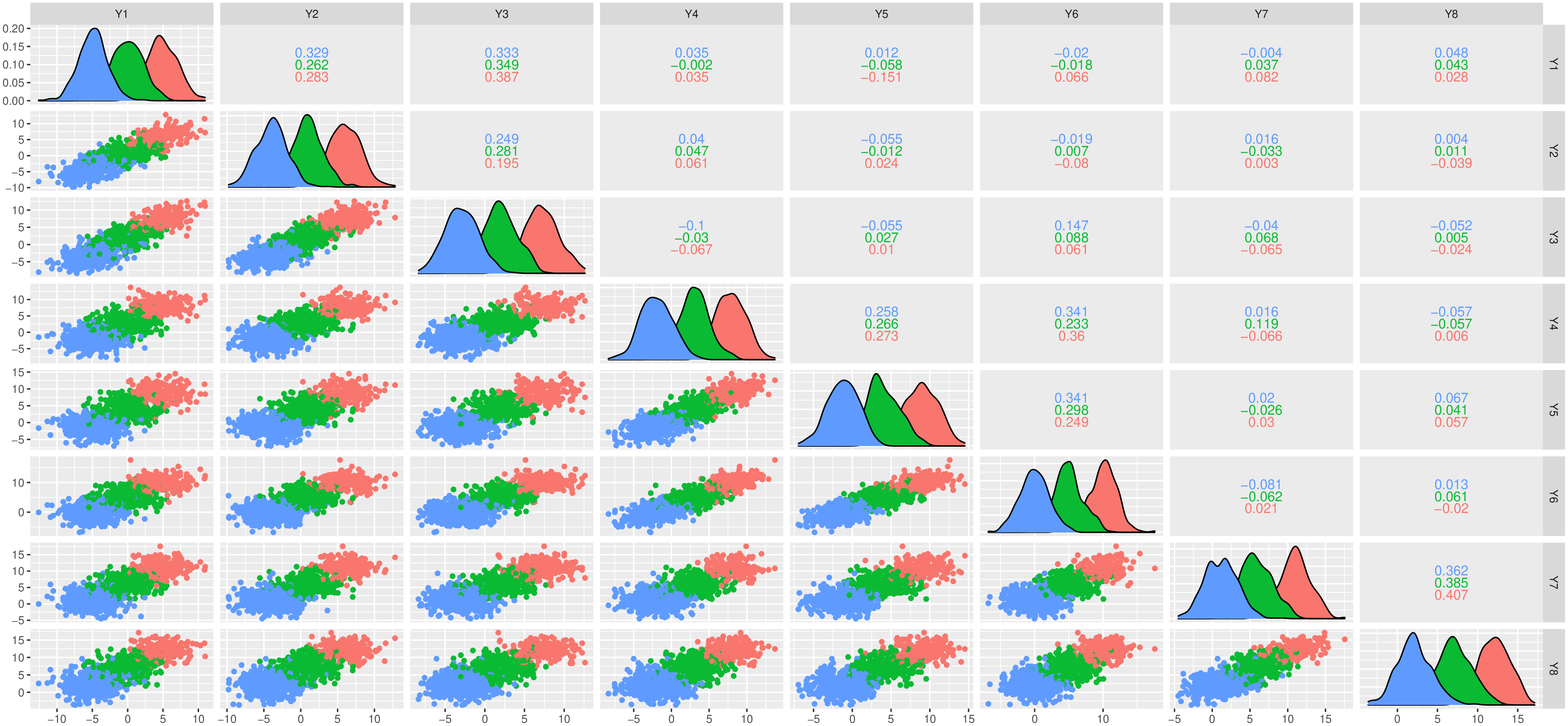}
	\caption{Scatterplot matrix for one of the hundred datasets for Simulation Study 1. The sub-plots above the diagonal sub-plots contains the sample correlation between the respective observed variables for each cluster.}\label{sim1fig}
\end{figure}

For the first simulation study, 100 datasets were generated from the most constrained CCCC model with K=3 and $Q=[3,3,3]$. The parameters used to generate the datasets are provided in Table \ref{sim1}. As can be seen in Figure \ref{sim1fig} (one of the hundred datasets), the clusters are not well-separated. In 99 out of the 100 datasets, the BIC selected the correct model with an average ARI of 0.98 (standard error of 0.01) and the estimated parameters are very close to the true parameters (summarized in Table \ref{sim1}).  
\begin{table}[!ht]
	\scriptsize
	\setlength{\tabcolsep}{5pt}
	\centering
	\caption{True parameters along with the averages and standard errors of the estimated values of the parameters from the 99 out of the 100 datasets where the correct CCCC model was selected.}\label{sim1}
	\begin{tabular}{@{\extracolsep{\fill}}cc|c}
		\\[-1.8ex]\hline
		\hline \\[-1.8ex]
		\multicolumn{1}{c}{}& \multicolumn{1}{c}{True parameters} &  \multicolumn{1}{c}{Average of estimated parameters}\\
		\multicolumn{1}{c}{}& \multicolumn{1}{c}{} &  \multicolumn{1}{c}{(standard errors)}\\
		
		\hline \\[-1.8ex]
		\multicolumn{1}{c}{}& \multicolumn{2}{c}{Component 1($n_1=500$)}\\
		\hline
		&&\\
		
		$\boldsymbol{\mu}_1$ &[-5, -4, -3, -2, -1, 0, 1, 2]&[-5.00, -4.00, -3.00, -1.99, -1.01, -0.01,  1.01,  2.01]\\
		&&(0.09, 0.10, 0.10, 0.09, 0.09, 0.09, 0.09, 0.09)\\
		$\pi_1$& 0.5 & 0.5 (0.01)\\
		\hline \\[-1.8ex]
		\multicolumn{1}{c}{}& \multicolumn{2}{c}{Component 2($n_2=300$)}\\
		\hline 
		&&\\
		$\boldsymbol{\mu}_2$ &[0, 1, 2, 3, 4, 5, 6, 7]&[0.00, 1.01, 2.01, 3.02, 4.02, 5.01, 6.03, 7.00]\\
		&&(0.12, 0.12, 0.11, 0.12, 0.12, 0.13, 0.13, 0.12)\\
		$\pi_2$& 0.3 & 0.3 (0.01)\\
		\hline \\[-1.8ex]
		\multicolumn{1}{c}{}& \multicolumn{2}{c}{Component 3($n_3=200$)}\\
		\hline
		&&\\
		$\boldsymbol{\mu}_3$ &[5, 6, 7, 8, 9, 10, 11, 12]&[5.00,  6.00 , 7.01 , 8.00 , 9.00 , 9.99, 10.98 ,11.99]\\
		&&(0.15, 0.19, 0.17, 0.16, 0.16, 0.16, 0.16, 0.14)\\
		$\pi_3$& 0.2 & 0.2 (0.01)\\
		\hline \\[-1.8ex]
		\multicolumn{1}{c}{}& \multicolumn{2}{c}{The common covariance matrix for all three components}\\
		\hline
		&&\\
		$\boldsymbol{\Sigma}$&$\left[\begin{matrix}
		4.5 & 2 & 2& 0 & 0 & 0& 0&0\\
		2 & 4.5 & 2& 0 & 0 & 0& 0&0\\
		2 & 2 & 4.5& 0 & 0 & 0& 0&0\\
		0 &0 &0& 4.5& 2 &2& 0& 0\\
		0 &0 &0& 2& 4.5 &2& 0& 0\\
		0 &0 &0& 2& 2 &4.5& 0& 0\\
		0 &0 &0& 0& 0 &0& 4.5 &2\\
		0& 0 &0& 0 &0& 0& 2& 4.5 
		\end{matrix}\right]$&$\left[\begin{matrix}
		4.47& 1.98& 1.98& 0& 0& 0& 0& 0\\
		1.98 &4.47& 1.98& 0& 0 &0& 0& 0\\
		1.98 &1.98& 4.47& 0& 0 &0& 0& 0\\
		0 &0 &0& 4.47& 1.98 &1.98& 0& 0\\
		0 &0 &0& 1.98& 4.47 &1.98& 0& 0\\
		0 &0 &0& 1.98& 1.98 &4.47& 0& 0\\
		0 &0 &0& 0& 0 &0& 4.47 &1.98\\
		0& 0 &0& 0 &0& 0& 1.98& 4.47 
		\end{matrix}\right]$\\
		&&\\
		sd($\boldsymbol{\Sigma}$)&&$\left(\begin{matrix}
		0.1& 0.09& 0.09& 0& 0& 0& 0& 0\\
		0.09 &0.1& 0.09& 0& 0 &0& 0& 0\\
		0.09&0.09& 0.1& 0& 0 &0& 0& 0\\
		0 &0 &0& 0.1& 0.09 &0.09& 0& 0\\
		0 &0 &0& 0.09& 0.1 &0.09& 0& 0\\
		0 &0 &0& 0.09& 0.09 &0.1& 0& 0\\
		0 &0 &0& 0& 0 &0& 0.1 &0.09\\
		0& 0 &0& 0 &0& 0& 0.09& 0.1 
		\end{matrix}\right)$\\
		\hline \\[-1.8ex]
	\end{tabular}
\end{table}

\subsection{Simulation Study 2}	
\begin{figure}[!htbp]
	\includegraphics[width=5.5in,height=3.5in]{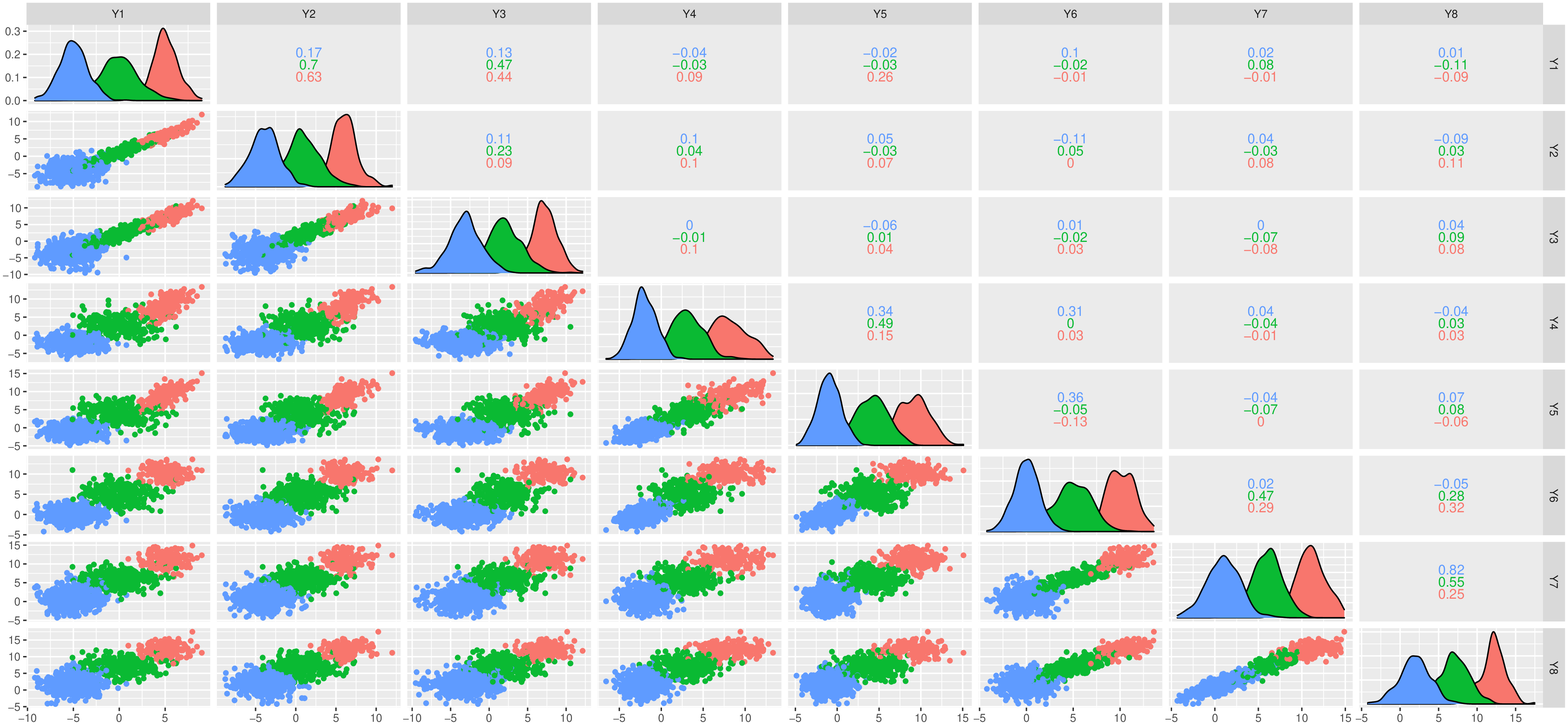}
	\centering\caption{Scatterplot matrix for one of the hundred datasets for Simulation Study 2. The sub-plots above the diagonal sub-plots contains the sample correlation between the respective observed variables for each cluster.}\label{sim2fig}
\end{figure}

For the second simulation, 100 datasets were generated from the completely unconstrained UUUU model with K=3 and $q=[3,3,2]$. The parameters used to generate the datasets are provided in Table \ref{sim2}. Figure \ref{sim2fig} provides a pairwise scatterplot matrix for one of the hundred datasets and again, the clusters are not well-separated. In 83 out of the 100 datasets, the BIC selected a three component model with some variations of $q$ and model types with an average ARI of 0.99 (standard error of 0.01) and in the remaining 17 datasets, a four component model was selected. In 44 out of those 83 datasets where the correct model (i.e., a three component UUUU model with $q=[3,3,2]$) was selected, the estimated parameters were close to the true parameters (see Table \ref{sim2}).

\newpage

{\scriptsize
	\begin{longtable}[!htbp]{cc|c}
		\caption{True parameters along with the averages and standard errors of the estimated values of the parameters from the  out of the 100 datasets where a $K=3$ model was selected.}\label{sim2}\\[2pt]
		\\[-1.8ex]\hline
		\hline \\[-1.8ex]
		\multicolumn{1}{c}{}& \multicolumn{1}{c}{True parameters} &  \multicolumn{1}{c}{Average of estimated parameters}\\
		\multicolumn{1}{c}{}& \multicolumn{1}{c}{} &  \multicolumn{1}{c}{(standard errors)}\\
		
		\hline \\[-1.8ex]
		\multicolumn{1}{c}{}& \multicolumn{2}{c}{Component 1($n_1=500$)}\\
		\hline
		&&\\
		$\pi_1$& 0.5 & 0.5 (0.01)\\[1ex]
		$\boldsymbol{\mu}_1$ &[-5, -4, -3, -2, -1, 0, 1, 2]&[-5.01, -4.00, -3.01, -2.00, -1.00, -0.01,  1.01 , 2.01]\\
		&&(0.06, 0.09, 0.10, 0.06, 0.06, 0.07, 0.08, 0.08)\\[1ex]
		$\boldsymbol{\Sigma}_1$&$\left[\begin{matrix}
		2.5 & 0.5 & 0.5 &   0  &  0  &  0 & 0&  0\\
		0.5  &3.5 & 0.5  &  0  &  0  &  0 & 0&  0\\
		0.5 & 0.5 & 4.5  &  0  &  0  &  0 & 0&  0\\
		0 & 0 & 0  &  2  &  1  &  1 & 0&  0\\
		0 & 0 & 0  &  1  &  2  &  1 & 0&  0\\
		0 & 0 & 0  &  1  &  1  &  2 & 0&  0\\
		0 & 0 & 0  &  0  &  0  &  0 & 3.5&  3\\
		0 & 0 & 0  &  0  &  0  &  0 & 3&  3.9\\
		\end{matrix}\right]$&$\left[\begin{matrix}
		2.50 & 0.50 & 0.50 & 0 & 0 & 0 & 0 & 0 \\ 
		0.50 & 3.51 & 0.50 & 0 & 0 & 0 & 0 & 0 \\ 
		0.50 & 0.50 & 4.49 & 0 & 0 & 0 & 0 & 0 \\ 
		0 & 0 & 0 & 2.00 & 1.01 & 1.01 & 0 & 0 \\ 
		0 & 0 & 0 & 1.01 & 2.00 & 1.01 & 0 & 0 \\ 
		0 & 0 & 0 & 1.01 & 1.01 & 2.00 & 0 & 0 \\ 
		0 & 0 & 0 & 0 & 0 & 0 & 3.48 & 2.99 \\ 
		0 & 0 & 0 & 0 & 0 & 0 & 2.99 & 3.89 \\ 
		\end{matrix}\right]$\\
		&&\\
		sd($\boldsymbol{\Sigma}_1$)&&$\left(\begin{matrix}
		0.18 & 0.12 & 0.12 & 0 & 0 & 0 & 0 & 0 \\ 
		0.12 & 0.20 & 0.12 & 0 & 0 & 0 & 0 & 0 \\ 
		0.12 & 0.12 & 0.26 & 0 & 0 & 0 & 0 & 0 \\ 
		0 & 0& 0 & 0.13 & 0.08 & 0.08 & 0 & 0 \\ 
		0 & 0 & 0 & 0.08 & 0.14 & 0.08 & 0 & 0 \\ 
		0 & 0 & 0 & 0.08 & 0.08 & 0.13 & 0 & 0 \\ 
		0 & 0 & 0 & 0 & 0 & 0 & 0.25 & 0.27 \\ 
		0 & 0& 0 & 0 & 0 & 0 & 0.27 & 0.33 \\ 
		\end{matrix}\right)$\\
		\hline \\[-1.8ex]
		\multicolumn{1}{c}{}& \multicolumn{2}{c}{Component 2($n_2=300$)}\\
		\hline 
		&&\\
		$\pi_2$& 0.3 & 0.3 (0.01)\\[1ex]
		$\boldsymbol{\mu}_2$ &[0, 1, 2, 3, 4, 5, 6, 7]&[-0.01,  1.00,  2.00,  3.02 , 4.01,  5.02 , 6.03 , 7.01]\\
		&&(0.11, 0.12, 0.12, 0.11, 0.12, 0.12, 0.11, 0.11)\\[1ex]
		$\boldsymbol{\Sigma}_2$&$\left[\begin{matrix}
		4.2 & 4 & 4 & 0 & 0 & 0 & 0 & 0 \\ 
		4 & 4.4 & 4 & 0 & 0 & 0 & 0 & 0 \\ 
		4 & 4 & 4.8 & 0 &0 & 0 & 0 & 0 \\ 
		0 & 0 & 0 & 4 & 2 & 0 & 0 & 0 \\ 
		0 & 0 & 0 & 2 & 4 & 0 & 0 & 0 \\ 
		0 & 0 & 0 & 0 & 0 & 4 & 3 & 3 \\ 
		0 & 0 & 0 & 0 & 0 & 3 & 3.5 & 3 \\ 
		0 & 0 & 0 & 0 & 0 & 3 & 3 & 4\\ 
		\end{matrix}\right]$&$\left[\begin{matrix}
		3.91 & 3.74 & 3.74 & 0 & 0 & 0 & 0 & 0 \\ 
		3.74 & 4.16 & 3.74 & 0 & 0 & 0 & 0 & 0 \\ 
		3.74 & 3.74 & 4.59 & 0 & 0 & 0 & 0 & 0 \\ 
		0 & 0 & 0 & 4.33 & 2.35 & 0 & 0 & 0 \\ 
		0 & 0 & 0 & 2.35 & 4.31 & 0 & 0 & 0 \\ 
		0 & 0 & 0 & 0 & 0 & 4.08 & 3.06 & 3.06 \\ 
		0 & 0 & 0 & 0 & 0 & 3.06 & 3.54 & 3.06 \\ 
		0 & 0 & 0 & 0 & 0 & 3.06 & 3.06 & 4.04 \\ 
		\end{matrix}\right]$\\
		&&\\
		sd($\boldsymbol{\Sigma}_2$)&&$\left(\begin{matrix}
		0.55 & 0.54 & 0.54 & 0 & 0 & 0 & 0 & 0 \\ 
		0.54 & 0.55 & 0.54 & 0 & 0 & 0 & 0 & 0 \\ 
		0.54 & 0.54 & 0.58 & 0 & 0 & 0 & 0 & 0 \\ 
		0 & 0 & 0 & 0.70 & 0.63 & 0 & 0 & 0 \\ 
		0 & 0 & 0 & 0.63 & 0.73 & 0 & 0 & 0 \\ 
		0 & 0 & 0 & 0 & 0 & 0.31 & 0.30 & 0.30 \\ 
		0 & 0 & 0 & 0 & 0 & 0.30 & 0.32 & 0.30 \\ 
		0 & 0 & 0 & 0 & 0 & 0.30 & 0.30 & 0.32 \\ 
		\end{matrix}\right)$\\
		\hline \\[-1.8ex]
		\multicolumn{1}{c}{}& \multicolumn{2}{c}{Component 3($n_3=200$)}\\
		\hline
		&&\\
		$\pi_3$& 0.2 & 0.2 (0.01)\\[1ex]
		$\boldsymbol{\mu}_3$ &[5, 6, 7, 8, 9, 10, 11, 12]&[5.01 , 6.01 , 7.02 , 8.01 , 9.01 , 9.99 ,10.97, 11.99]\\
		&&(0.10, 0.13, 0.14, 0.16, 0.15, 0.11, 0.14, 0.10)\\[1ex]
		$\boldsymbol{\Sigma}_3$&$\left[\begin{matrix}
		2.1 & 2 & 2 & 2 & 2 & 0 & 0 & 0 \\ 
		2 & 2.5 & 2 & 2 & 2 & 0 & 0 & 0 \\ 
		2 & 2 & 3 & 2 & 2 & 0 & 0 & 0 \\ 
		2 & 2 & 2 & 5 & 2 & 0 & 0 & 0 \\ 
		2 & 2 & 2 & 2 & 4 & 0 & 0 & 0 \\ 
		0 & 0 & 0 & 0 & 0 & 2 & 1 & 1 \\ 
		0 & 0 & 0 & 0 & 0 & 1 & 3 & 1 \\ 
		0 & 0 & 0 & 0 & 0 & 1 & 1 & 2.5 \\ 
		\end{matrix}\right]$&$\left[\begin{matrix}
		1.99 & 1.89 & 1.89 & 1.89 & 1.89 & 0 & 0 & 0 \\ 
		1.89 & 2.40 & 1.89 & 1.89 & 1.89 & 0 & 0 & 0 \\ 
		1.89 & 1.89 & 2.90 & 1.89 & 1.89 & 0 & 0 & 0 \\ 
		1.89 & 1.89 & 1.89 & 4.89 & 1.89 & 0 & 0 & 0 \\ 
		1.89 & 1.89 & 1.89 & 1.89 & 3.94 & 0 & 0 & 0 \\ 
		0 & 0 & 0 & 0 & 0 & 2.13 & 1.15 & 1.15 \\ 
		0 & 0 & 0 & 0 & 0 & 1.15 & 3.16 & 1.15 \\ 
		0 & 0 & 0 & 0 & 0 & 1.15 & 1.15 & 2.62 \\ 
		\end{matrix}\right]$\\
		&&\\
		sd($\boldsymbol{\Sigma}_3$)&&$\left(\begin{matrix}
		0.26 & 0.28 & 0.28 & 0.28 & 0.28 & 0 & 0 & 0 \\ 
		0.28 & 0.31 & 0.28 & 0.28 & 0.28 & 0 & 0 & 0 \\ 
		0.28 & 0.28 & 0.32 & 0.28 & 0.28 & 0 & 0 & 0 \\ 
		0.28 & 0.28 & 0.28 & 0.43 & 0.28 & 0 & 0 & 0 \\ 
		0.28 & 0.28 & 0.28 & 0.28 & 0.38 & 0 & 0 & 0 \\ 
		0 & 0 & 0 & 0 & 0 & 0.26 & 0.28 & 0.28 \\ 
		0 & 0 & 0 & 0 & 0 & 0.28 & 0.37 & 0.28 \\ 
		0 & 0 & 0 & 0 & 0 & 0.28 & 0.28 & 0.32 \\ 
		\end{matrix}\right)$\\	
		\hline \\[-1.8ex]
	\end{longtable}	
}

\subsection{Real data analysis}
We applied our method to 3 datasets:
\begin{enumerate}
	\item \texttt{Alon} data \citep{Alon:1999dy} contains the gene expression measurements of 6500 genes using an Affymetrix oligonucleotide Hum6000 array of 62 samples (40 tumor samples, 22 normal samples) from colon-cancer patients. We started with the preprocessed version of the data from \cite{mcnicholas2010} that comprised of 461 genes. As $p>>n$ and  our algorithm is currently not designed for high dimensional data, to reduce the dimensionality, a $t$-test followed by false discovery rate (FDR) threshold of 0.1\% was used that yielded in 22 differentially expressed genes. Hence, the resulting dimensionality of the dataset to the sample size ratio (i.e. $\frac{p}{n}\approx 0.55$).
	
	\item \texttt{Golub} data \citep{Golub99molecularclassification} contains gene expression values of 7129 genes from 72 samples:  47 patients with acute lymphoblastic leukemia (ALL) and 25 patients with acute myeloid leukemia (AML). We started with the preprocessed version of the data from \cite{mcnicholas2010} that comprised of 2030 genes. Again, as $p>>n$ and our algorithm is currently not designed for high dimensional data, to reduce the dimensionality, we select top 40 most differentially expressed genes by setting the FDR threshold as 0.00001\%. Hence, the resulting dimensionality of the dataset to the sample size ratio (i.e. $\frac{p}{n}\approx 0.55$).

	\item \texttt{Wine} data available in the {\sf R} package \texttt{rattle}\citep{rattle}, contains information on the 13 different attributes from the chemical analysis of wines grown in specific areas of Italy. The dataset comprises of 178 samples of wine that can be categorized into three types: ``Barolo", ``Grignolino", and ``Barbera". Since the sample size $n>>p$, we use all 13 variables here. Since the chemical measurements are in different scales, the dataset was scaled before running biclustering methods. 
\end{enumerate}

For each dataset, we ran all of the 16 proposed models for $K = 1\ldots4$ and $q_k = 1\ldots8$. For comparison, we show the clustering results of the following biclustering approaches applied to the above real datasets:
\begin{enumerate}
	\item U-OSGaBi family (Unsupervised version of OSGaBi family): Here, we run the approach proposed by \cite{wong2017two}. In \cite{wong2017two}, the authors did one-way supervision (assuming observation's memberships as unknowns and variable's group memberships as knowns) however in our analysis we perform unsupervised clustering for both rows and columns. We run all 8 models by \cite{wong2017two} for $K = 1\ldots4$ and $q = 1\ldots8$ by \cite{wong2017two}. Note that these models can be obtained as a special case of our proposed models by imposing the restriction that $q_1=q_2=\ldots=q_k=q$ and $\bT_{q_k}=\bI$. 	
	\item \underline{Block-cluster}: Here, we also run the biclustering models proposed by \cite{blockcluster} for continuous data. This model utilizes mixtures of univariate Gaussian distributions. All four models obtained via imposition of constraints on the mixing proportions and variances to be equal or different across groups were run using the {\sf R} package ``blockcluster"\citep{blockcluster}.

\end{enumerate}
The performance of all three methods on the real datasets are summarized in Table \ref{compare}. Our proposed model outperforms both U-OSGaBi family and ``block-cluster" method on \texttt{Alon} data and \texttt{Wine}. However, both ``block-cluster" method and our proposed method provide the same clustering performance on the \texttt{Golub} data. It is interesting to note that on the \texttt{Wine} data, 
the model selected by our approach is UUCU model and the model selected from U-OSGaBi family is UCU model. Both models have the same constraints for $\bB_k$ and $\bD$, however, in U-OSGaBi family, there is an additional restriction that $\bT_{q_k}=\bI$. Removing the restriction that $\bT_{q_k}=\bI$ in our approach gives a substantial increase in the ARI (i.e from 0.74 to 0.93). Additionally, on the \texttt{Golub} dataset, fitting U-OSGaBi family results in the selection of $q=3$ and our proposed algorithm chooses $q=6$, both with the same constrain for $\bB$. The CUU model selected for U-OSGaBi family has a constrained $\bB$ matrix and fixed $\bT=\bI$ whereas our proposed approach also selects a model with constrained $\bB$ matrix but a group-specific anisotropic matrix $\bT_k$. However, the ARI from our proposed method (i.e., ARI=0.94) is much higher than the ARI from U-OSGaBi family (i.e., ARI=0.69).

\begin{figure}[!htbp]
	\includegraphics[width=1\textwidth]{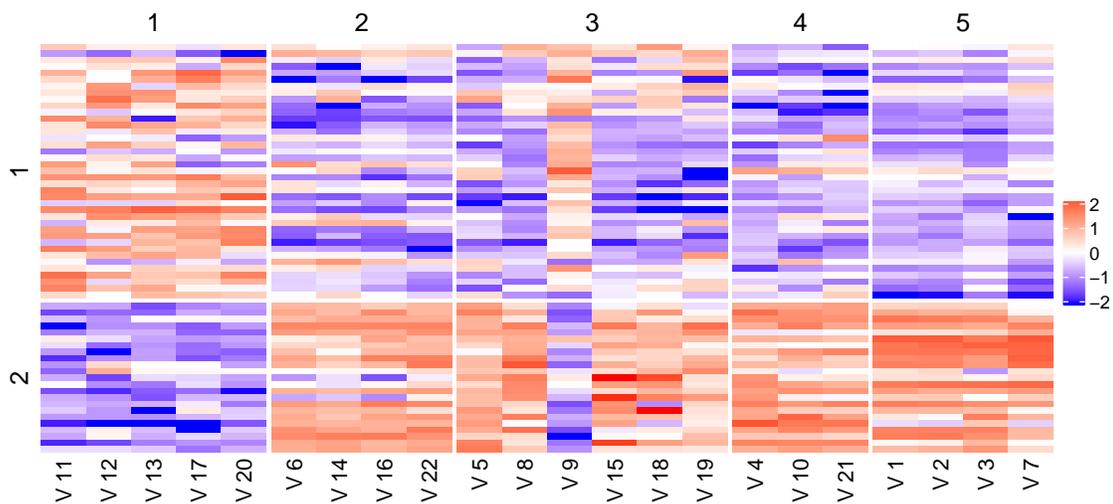}
	\caption{Heatmap of the observations from the \texttt{Alon} data.}\label{alon_obs}
\end{figure}

Also, notice that for \texttt{Wine} data, our proposed model selects different values for $q$ for different groups. Hence, this improvement in the clustering performance could be due to removal of the restriction that $\bT_{q_k}=\bI$ for all $k=1,\ldots,K$, removal of the restriction $q_1=q_2=\ldots=q_K=q$, or both. 
Here, we will use \texttt{Alon} data for detailed illustration. While ARI can be used for evaluating the agreement of the row cluster membership with a reference class indicator variable, a heatmap of the observations is typically used to visualize the bicluster structure. Figure \ref{alon_obs} shows that our proposed method is able to recover the underlying bicluster structure fairly well.

\begin{figure}[!htbp]
	\begin{subfigure}{.5\textwidth}
		\includegraphics[width=1\textwidth]{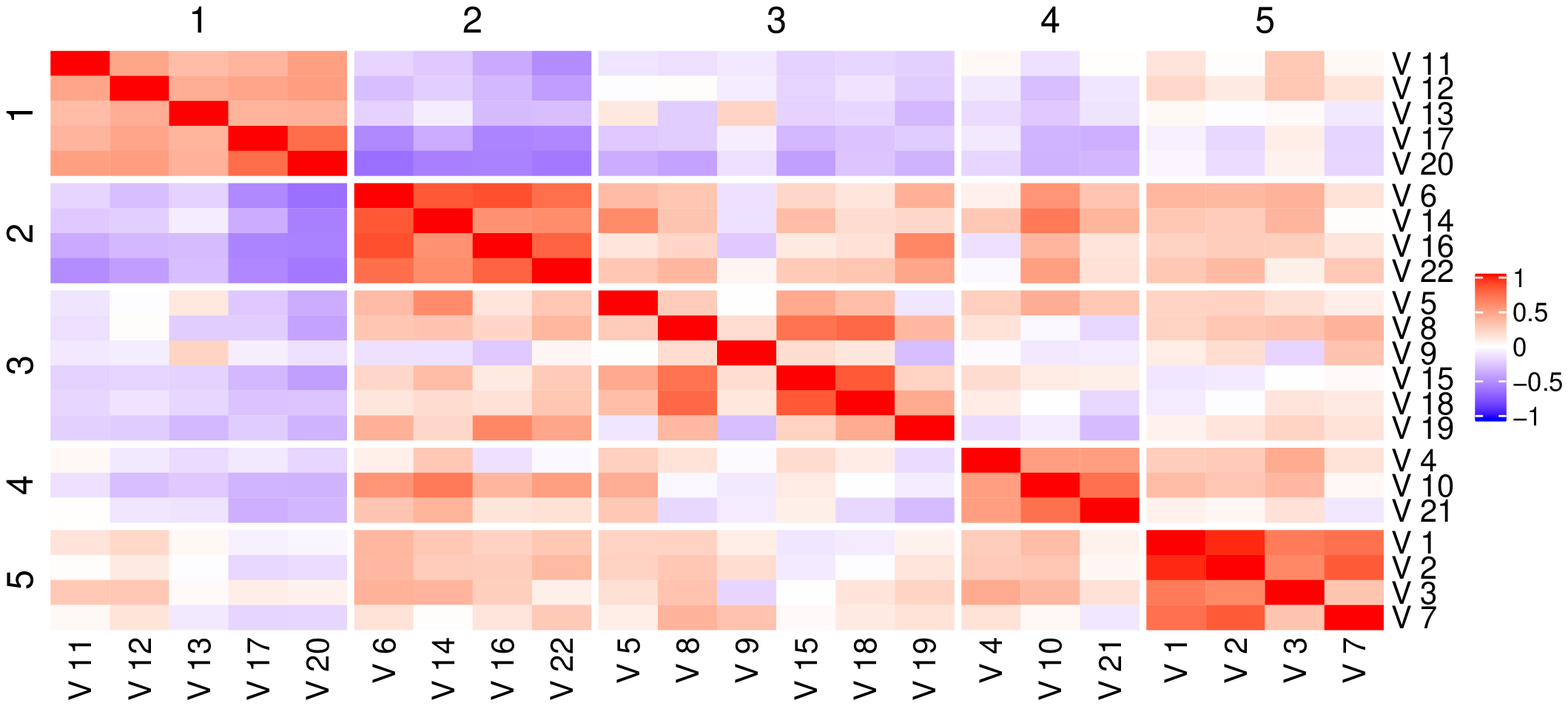}
		\includegraphics[width=1\textwidth]{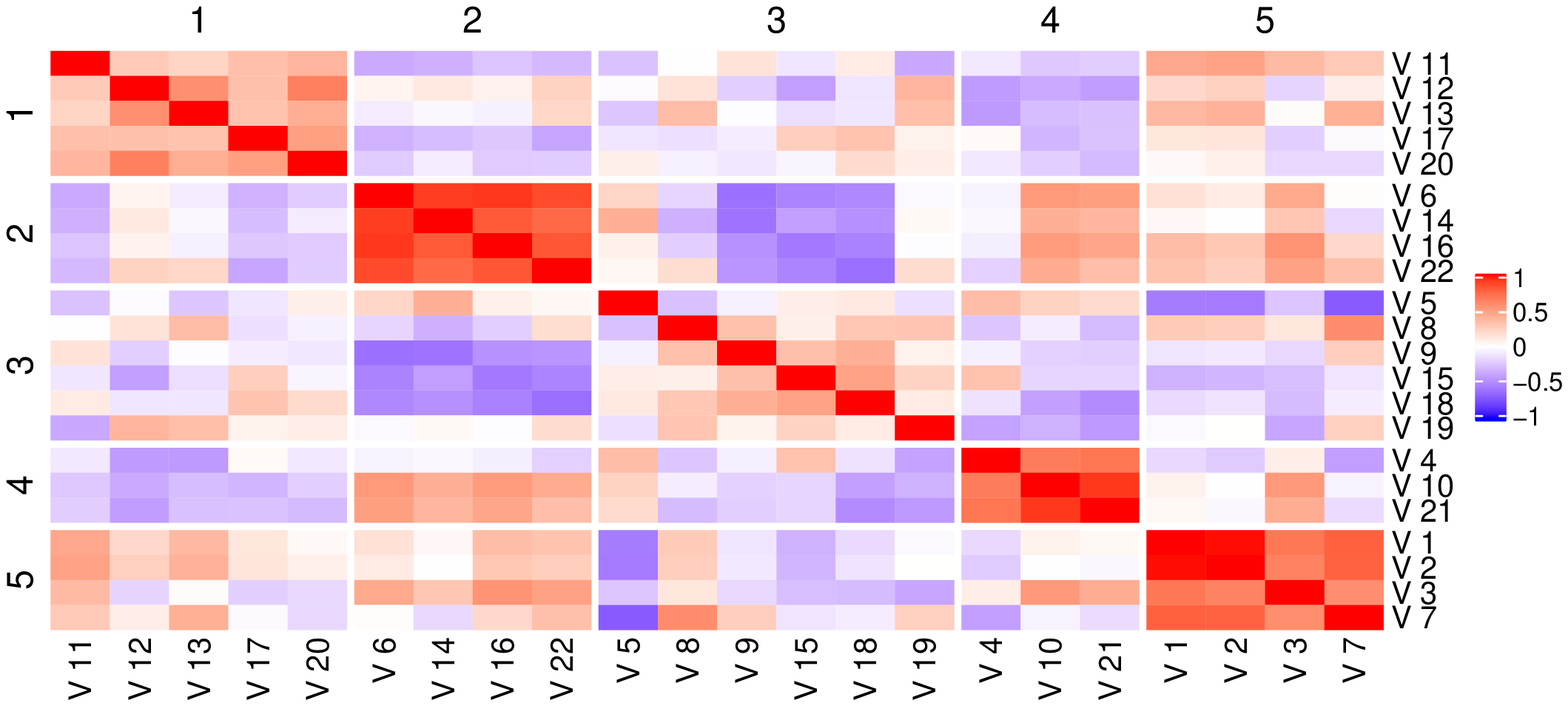}
		\caption{Observed correlation structures.}
	\end{subfigure}
	\begin{subfigure}{.5\textwidth}
		\includegraphics[width=1\textwidth]{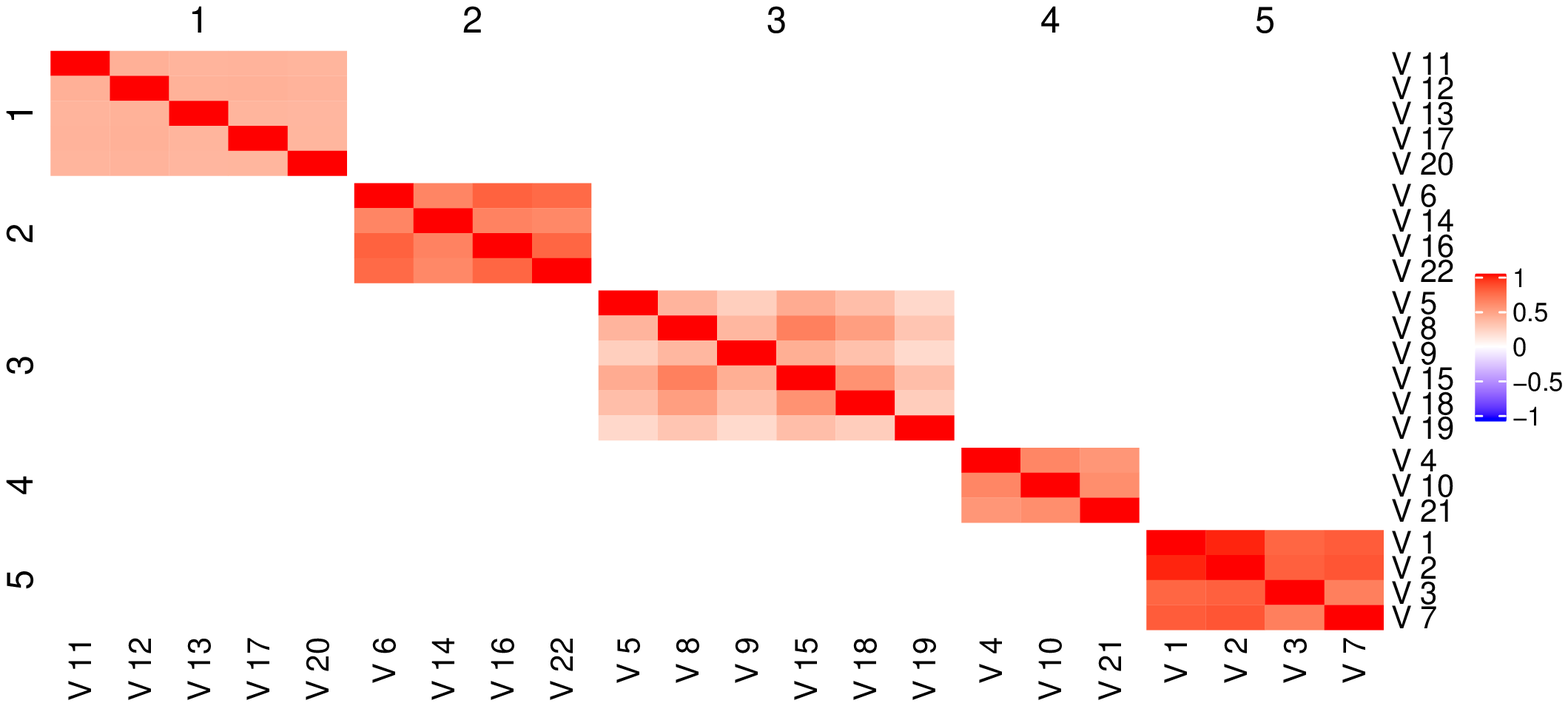}
		\includegraphics[width=1\textwidth]{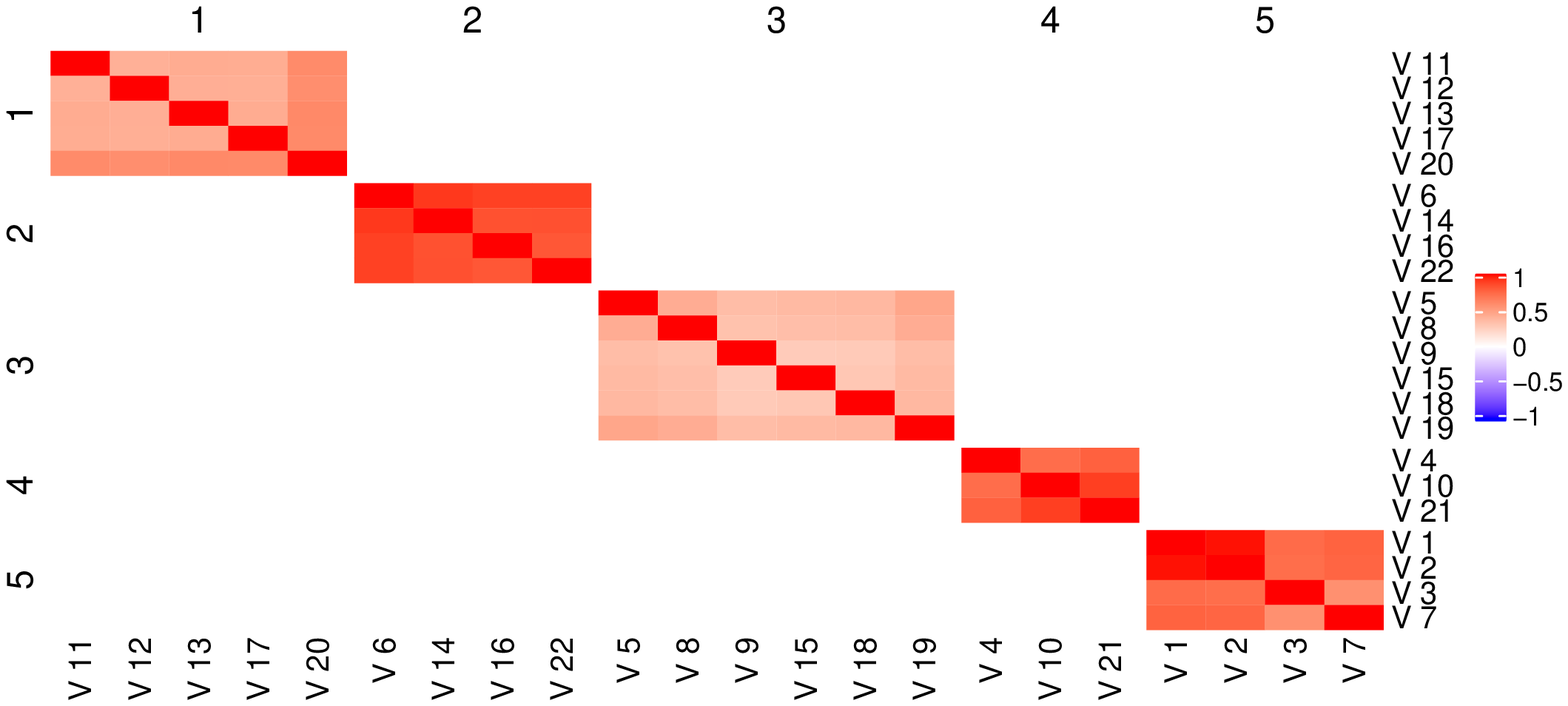}
		\caption{Recovered correlation structures.}
	\end{subfigure}
	\caption{Heatmap of the cluster-specific correlation structures in the two row clusters of the \texttt{Alon} data.}\label{alon_cor}
\end{figure}

We also visualize component correlation matrices to gain an insight into the observed column clusters. As evident from the heatmap of the observed and estimated covariance matrices in Figure \ref{alon_cor}, variables that are highly correlated are together in the same column clusters.

\begin{table}[!htbp]
	\centering
	\caption{Summary of the clustering performances by the best model selected using BIC for all three approaches.}\label{compare}
	\begin{tabular}{@{\extracolsep{\fill}}ccccccc}
		\\[-1.8ex]\hline
		\hline \\[-1.8ex]
		
		Data&True \# of Classes&Approach&Model selected&K& $\mathbf{q}_k$ & ARI\\
		\hline
		&&&&&\\
		\texttt{Alon}&2&Proposed&CCUU&2&$[5,5]$&\textbf{0.69}\\
		&&U-OSGaBi family&CCU &2&[5,5]&0.64\\
		&&Block-cluster&``pik\_rhol\_sigma2"&4&$[2,2]$&0.33\\
		\hline
		&&&&&\\
		\texttt{Golub}&2&Proposed&CUUU&2&$[6,6]$&\textbf{0.94}\\	
		&&U-OSGaBi family& CUU &2&[3,3]&0.69\\	
		&&Block-cluster&``pi\_rho\_sigma2kl"&2&$[4,4]$&\textbf{0.94}\\	
		\hline
		&&&&&\\	
		\texttt{Wine}&3&Proposed&UUCU&3&$[5,7,5]$&\textbf{0.93}\\
		&&U-OSGaBi family&UCU&3&[5,5,5]&0.74\\
		&&Block-cluster&``pik\_rhol\_sigma2kl"&3&$[7,7,7]$&0.88\\
		\hline
		\hline \\[-1.8ex]
	\end{tabular}
\end{table}

\section{Conclusion}\label{conc}
In this paper, we propose an extended family of 16 models for model-based for biclustering. Parsimony is introduced in two ways. First, as the factor loading matrix $\bB$ matrix is binary row-stochastic, the resulting covariance matrix is block-diagonal. Secondly, we  also impose constraints on elements of the covariance matrices that results in a family of models with varying number of parameters. Our proposed method builds on the work by \cite{Martella2008BiclusteringOG} and \cite{wong2017two} which utilized a factor analyzer structure for developing a model-based biclustering framework. However, those works restricted the covariance matrix of the latent variable to be an identity matrix and assumed that the number of latent variables was the same for all components. The restriction on the covariance matrix of the latent variable imposes a restriction on the structure on the block diagonal part of the covariance matrix, therefore only allowing for high variance-low covariance or low variance-high covariance structure. Here, we propose a modified factor analyzer structure that assumes that the covariance of the latent variable is a diagonal matrix, and therefore, is able to recover a wide variety of covariance structures. Our simulations demonstrate that good parameter recovery. Additionally, we also allow different components to have different $q$, and therefore, allowing different grouping of the variables in different clusters. Using simulated data, we demonstrate that these models give a good clustering performance and can recover the underlying covariance structure. In real datasets, we show through comparison with the method by \cite{wong2017two} that easing the restrictions on $\bT$ and on the number of latent variables can provide substantial improvement in the clustering performance. We also compared our approach to the block-cluster method by \cite{blockcluster} and show that our proposed method provides a competitive performance.

Although our proposed models can capture an extended range of the covariance structure compared to \cite{Martella2008BiclusteringOG} and \cite{wong2017two}, it still can only allow positive correlations within the column clusters. However, typically in biology, it may be of interest to group variables based on the magnitude of the correlation regardless of the sign of the correlation. For example, suppose a particular pathway plays a crucial role in a tumor development and consequently, genes involved in that pathways show changes in their expression levels. Some gene may be up-regulated while others may be down-regulated, and hence, these genes will be divided into multiple column clusters which will lead over estimation of the number of column clusters. Additionally, each column cluster will only provide a partial and incomplete view of the pathway's involvement. Furthermore, investigation of approaches for efficient update of $\bB_k$ is warranted. While the current approach of updating it row by row provided satisfactory clustering performance, it is not computationally efficient and can sometime miss the true underlying structure. Our proposed approach only allows for variables to be in one column cluster whereas from a practical viewpoint, introducing soft margin and allowing variables to be involved in more than one column clusters might be informative.



\section*{Appendix}
\subsection*{A: Estimation for 16 models in the family}\label{appendix: family}
Here, we provide parameter estimates for the components of the covariance matrices for all 16 proposed models. Recall the following:
\[
\mathbf{S}_k=\frac{\sum_{i=1}^{n}\hat{z}_{ik}(\by_i-\boldsymbol{\mu}_k)(\by_i-\boldsymbol{\mu}_k)^T}{n_k}, \quad n_k=\sum_{i=1}^{n}\hat{z}_{ik}, \text{and} \quad \boldsymbol{\theta}_k=\frac{\sum_{i=1}^n \hat{z}_{ik}E(\bU_{ik}\bU_{ik}^T|\by_i)}{n_k}.
\]

\begin{enumerate}
	\item UUUU model: Here, we assume no constraints on $\bB_k, \bT_{q_k}$, and $\bD_k$.
	Details on the parameter estimates are provided in Section \ref{PE}.	
	\item UUUC model: Here, we assume $\bD_k=d_k\mathbf{I}_p$, and no constraints on $\bB_k$ and $\bT_{q_k}$. 
	The estimates of  $\bB_k$ and  $\bT_{q_k}$ are the same as UUUU model. Simply replace $\bD_k$ with $d_k\mathbf{I}_p$ in equation \ref{UUUQ2}, then taking derivative respect to $d_k$ yeilds: 
	\[
	\hat{d_k}=\frac{1}{p}\text{tr}\left\{\mathbf{S}_k-2\hat{\bB}_k\hat{\bT}_{q_k}\hat{\bB}_k^T\left(\hat{\bB}_k\hat{\bT}_{q_k}\hat{\bB}_k^T+\hat{\bD}_k^{(t)}\right)^{-1}\mathbf{S}_k+\hat{\bB}_k\boldsymbol{\theta}_k\hat{\bB}_k^T\right\}.
	\]
	
	\item UUCU model: Here, we assume $\bD_k=\bD$, and no constraints for $\bB_k$ and $\bT_{q_k}$. The estimates of  $\bB_k$ and  $\bT_{q_k}$ are the same as UUUU model. The estimate for $\bD$ is
	\[
	\hat{\bD}=\sum_{k=1}^{K}\hat{\pi}_k\text{diag}\left\{\mathbf{S}_k-2\hat{\bB}_k\hat{\bT}_{q_k}\hat{\bB}_k^T\left(\hat{\bB}_k\hat{\bT}_{q_k}\hat{\bB}_k^T+\hat{\bD}^{(t)}\right)^{-1}\mathbf{S}_k+\hat{\bB}_k\boldsymbol{\theta}_k\hat{\bB}_k^T\right\}.
	\]
	
	\item UUCC model:  Here, we assume $\bD_k=d\mathbf{I}_p$, and no constraints for $\bB_k$ and $\bT_{q_k}$. The estimates of  $\bB_k$ and  $\bT_{q_k}$ are the same as UUUU model. The estimate for $d$ is
	\[
	\hat{d}=\sum_{k=1}^{K}\frac{\hat{\pi}_k}{p}~\text{tr}\left\{\mathbf{S}_k-2\hat{\bB}_k\hat{\bT}_{q_k}\hat{\bB}_k^T\left(\hat{\bB}_k\hat{\bT}_{q_k}\hat{\bB}_k^T+\hat{\bD}^{(t)}\right)^{-1}\mathbf{S}_k+\hat{\bB}_k\boldsymbol{\theta}_k\hat{\bB}_k^T\right\}.
	\]
	\item UCUU model: Here, we assume $\bT_{q_k}=\bT$, and no constraints for $\bB_k$ and $\bD_k$. 
	The estimates of  $\bB_k$ and  $\bD_k$ are the same as UUUU model and the estimate for $\bT$ is
	\[
	\hat{\bT}=\sum_{k=1}^{K}\hat{\pi}_k~\text{diag}(\boldsymbol{\theta}_k).
	\]
	
	\item UCUC model: Here, we assume $\bT_{q_k}=\bT, \bD_k=d_k\mathbf{I}_p$, and no constraint for $\bB_k$. The estimate of  $\bB_k$ is the same as UUUU model and the estimates are of $\bT$ and $d_k$ are
	\begin{align*}
	\hat{\bT}&=\sum_{k=1}^{K}\hat{\pi}_k\text{diag}(\boldsymbol{\theta}_k),\\
	\hat{d}_k&=\frac{1}{p}\text{tr}\left\{\mathbf{S}_k-2\hat{\bB}_k\hat{\bT}\hat{\bB}_k^T\left(\hat{\bB}_k\hat{\bT}\hat{\bB}_k^T+\hat{\bD}_k^{(t)}\right)^{-1}\mathbf{S}_k+\hat{\bB}_k\boldsymbol{\theta}_k\hat{\bB}_k^T\right\}.
	\end{align*}

	\item UCCU model: Here, we assume $\bT_{q_k}=\bT$ and $\bD_k=\bD$, and no constraint for $\bB_k$. The estimate of  $\bB_k$ is the same as UUUU model and the estimates are of $\bT$ and $\bD$ are
	\begin{align*}
	\hat{\bT}&=\sum_{k=1}^{K}\hat{\pi}_k~\text{diag}(\boldsymbol{\theta}_k),\\
	\hat{\bD}&=\sum_{k=1}^{K}\hat{\pi}_k~\text{diag}\left\{\mathbf{S}_k-2\hat{\bB}_k\hat{\bT}\hat{\bB}_k^T\left(\hat{\bB}_k\hat{\bT}\hat{\bB}_k^T+\hat{\bD}^{(t)}\right)^{-1}\mathbf{S}_k+\hat{\bB}_k\boldsymbol{\theta}_k\hat{\bB}_k^T\right\}.
	\end{align*}

	\item UCCC model: Here, we assume $\bT_{q_k}=\bT$ and $\bD_k=d\mathbf{I}_p$, and no constraint for $\bB_k$. The estimate of  $\bB_k$ is the same as UUUU model and the estimates are of $\bT$ and $d$ are
	\begin{align*}
	\hat{\bT}&=\sum_{k=1}^{K}\hat{\pi}_k~\text{diag}(\boldsymbol{\theta}_k),\\
	\hat{d}&=\sum_{k=1}^{K}\frac{\hat{\pi}_k}{p}~\text{tr}\left\{\mathbf{S}_k-2\hat{\bB}_k\hat{\bT}_{q_k}\hat{\bB}_k^T\left(\hat{\bB}_k\hat{\bT}_{q_k}\hat{\bB}_k^T+\hat{\bD}^{(t)}\right)^{-1}\mathbf{S}_k+\hat{\bB}_k\boldsymbol{\theta}_k\hat{\bB}_k^T\right\}.
	\end{align*}
	
	\item CUUU model: Here, we assume $\bB_k=\bB$, and no constraints for $\bT_{Q_k}$ and  $\bD_k$. The parameter estimates for $\bT_{Q_k}$ and  $\bD_k$ are exactly the same as UUUU model. For parameter estimate for $\bB$, we define
	$Q^*_2$ as
	\[
	Q^*_2=\sum_{k=1}^{K}n_k(tr\{\bD_k^{-1}\bB\bT_{q_k}\bB^T(\bB\bT_{q_k}\bB^T+\bD_k)^{-1}\boldsymbol{S}_k\}-\frac{1}{2}tr\{\bB^T\bD_k^{-1}\bB\boldsymbol{\theta}_k\}).
	\]
	When estimate $\bB$, we choose $\bB[i,j]=1$. when $\bB$ maximize $Q_2^*$, with constrain $\sum_{j=1}^{q_k}\bB[i,j]=1$ .
	
	\item CUUC model: Here, we assume $\bB_k=\bB, \bD_k=d_k\mathbf{I}_p$, and no constraint for $\bT_{q_k}$. Estimation of $\bB_k$ are exactly same as CUUU model and estimation of $d_k$ and $\bT_{q_k}$ are the same as UUUC model. 
	
	\item CUCU model: Here, we assume $\bB_k=\bB, \bD_k=\bD$, and no constraint for $\bT_{q_k}$. Estimation of $\bB_k$ are exactly same as CUUU model and estimation of $\bD$ and $\bT_{q_k}$ are the same as UUCU model.

	\item CUCC model: Here, we assume $\bB_k=\bB, \bD_k=d\mathbf{I}_p$, and no constrain for $\bT_{q_k}$. Estimation of $\bB_k$ are exactly same as CUUU model and the estimation of $d$ and $\bT_{q_k}$ are the same as UUCC model.
	
	\item CCUU model: Here, we assume $\bB_k=\bB, \bT_{q_k}=\bT$, and no constrain for $\bD_k$. Estimation of $\bB_k$ are exactly same as CUUU model and the estimation of $\bD_k$ and $\bT$ are the same as UCUU model.
	
	\item CCUC model: Here, we assume $\bB_k=\bB$, $\bT_{q_k}=\bT$, and $\bD_k=d_k\mathbf{I}_p$. Estimation of $\bB_k$ are exactly same as CUUU model and the estimation of $d_k$ and $\bT$ are the same as UCUC model. 
	
	\item CCCU model: Here, we assume $\bB_k=\bB$, $\bT_{q_k}=\bT$, and $\bD_k=\bD$. Estimation of $\bB_k$ are exactly same as CUUU model and the estimation of $\bD$ and $\bT$ are the same as UCCU model.  
	
	\item CCCC model: Here, we assume $\bB_k=\bB$, $\bT_{q_k}=\bT$, and $\bD_k=d\mathbf{I}_p$. Estimation of $\bB_k$ are exactly same as CUUU model and the estimation of $d$ and $\bT$ are the same as UCCC model  
	
\end{enumerate}

\subsection*{B: Special cases}\label{appendix: special case}
Here, we show that the family of models proposed by \cite{Martella2008BiclusteringOG} and \cite{wong2017two} can be obtained as special cases of our proposed models. 
\subsubsection*{Models by  \cite{Martella2008BiclusteringOG}}
If we assume the constraints that $\bT_{q_k}=\mathbf{I}$, the four models proposed by \cite{Martella2008BiclusteringOG} can be obtained as following:
\begin{table}[!htbp]
	\scriptsize
	\centering
	\caption{Models by  \cite{Martella2008BiclusteringOG}}\label{Martella}
	\begin{tabular}{cccccc}
		\\[-1.8ex]\hline
		\hline \\[-1.8ex]
		\multicolumn{1}{c}{ Models from \cite{Martella2008BiclusteringOG}}&\multicolumn{4}{c}{Equivalent constraints for proposed models}&\\
		&  \multicolumn{1}{c}{$\bB_k$}&\multicolumn{1}{c}{$\bT_{q_k}$}&\multicolumn{2}{c}{$\bD_k$}&\multicolumn{1}{c}{Total \# of parameters}\\
		\cline{3-3}\cline{4-5}
		&&Group&Group&Diagonal&\\
		&$\bB_k=\bB$&$\bT_{q_k}=\mathbf{I}$&$\bD_k=\bD$& $\bD_k=d_k \mathbf{I}$&\\
		\hline\\[-1.8ex]
		UU & U&I&U&U&$3pK+K-1$\\
		UC&U&I&C&U&$2pK+p+K-1$\\
		CU&C&I&U&U&$2pK+p+K-1$\\
		CC&C&I&C&U&$2p+K-1+pK$\\
		\hline \\[-1.8ex]
	\end{tabular}
\end{table}

\subsubsection*{Models by  \cite{wong2017two}}
If we further allow constraint that $\bD_k=d_k\mathbf{I}_p$ and $q_1=q_2=\ldots=q_K=q$, then we have the eight models by \cite{wong2017two}.
\begin{table}[!ht]
	\scriptsize
	\centering
	\caption{Models by  \cite{wong2017two}}\label{Wong}
	\begin{tabular}{@{\extracolsep{5pt}}cccccc}
		\\[-1.8ex]\hline
		\hline \\[-1.8ex]
		\multicolumn{1}{c}{ Models from \cite{wong2017two}}&\multicolumn{4}{c}{Equivalent constraints for proposed models}&\\
		&  \multicolumn{1}{c}{$\bB_k$}&\multicolumn{1}{c}{$\bT_{q_k}$}&\multicolumn{2}{c}{$\bD_k$}&\multicolumn{1}{c}{Total \# of parameters}\\
		\cline{3-3}\cline{4-5}
		&&Group&Group&Diagonal&\\
		&$\bB_k=\bB$&$\bT_{q_k}=\mathbf{I}$&$\bD_k=\bD$& $\bD_k=d_k \mathbf{I}$&\\
		\hline\\[-1.8ex]
		UUU & U&I&U&U&$3pK+K-1$\\
		UCU&U&I&C&U&$2pK+p+K-1$\\
		CUU&C&I&U&U&$2pK+p+K-1$\\
		CCU&C&I&C&U&$2p+K-1+pK$\\
		UUC & U&I&U&C&$2pK+2K-1$\\
		UCC&U&I&C&C&$2pK+K$\\
		CUC&C&I&U&C&$pK+p+2K-1$\\
		CCC&C&I&C&C&$p+K+pK$\\
		\hline \\[-1.8ex]
	\end{tabular}
\end{table}
\end{document}